\documentstyle[amssymb,aps,multicol,prl,epsf]{revtex}
\begin{document}

\title{Quantum fluctuations and the c-axis optical conductivity of High-$T_c$
Superconductors}

\author{L. B. Ioffe and A. J. Millis \\
Center for Materials Theory\\
Department of Physics and Astronomy\\ 
Rutgers University
Piscataway, NJ 08854}

\maketitle

\begin{abstract}
A theory of the frequency dependence of the interplane conductivity of a
strongly anisotropic superconductor is presented. The form of the
conductivity is shown to be a sensitive probe of the strength of quantum
and thermal fluctuations of the phase of the superconducting order parameter. The
temperature dependence of the superfluid stiffness and of the form of
the absorbtion at frequencies of the order of twice the superconducting gap
is shown to depend on the
interplay between superconducting pairing, phase coherence
and the mechanism by which electrons are
scattered. Measurements of the c-axis conductivity of high-$T_c$ 
superconductors are interpreted in terms of the theory.
\end{abstract}

\begin{multicols}{2}

\section{Introduction}

One of the important unresolved questions in high-$T_{c}$ superconductivity
is the strength of quantum fluctuations in the superconducting ground state.
In this paper we show that the frequency dependence of the c-axis
conductivity is a useful probe of the strength of these fluctuations. Our
results apply, with minor modifications which are indicated at appropriate
points below, to any sufficiently anisotropic
superconductor and therefore may
also be useful for interpreting data on layered
organic superconductors and on the 'spin-ladder' compounds.
Another important issue concerns the nature of the electronic
states in the $CuO_2$ planes of high-$T_c$ materials.  We will show
that the c-axis conductivity can be used to extract information
about these states.

In a previous paper \cite{Ioffe99} we presented a theory for the c-axis
optical spectral weight of layered superconducting systems. We showed, among
other things, that the strength of the quantum fluctuations could be
inferred from the ratio of the spectral weight in the c-axis superfluid
response to the spectral weight lost from the c-axis conductivity as the
temperature is decreased from high $T$ to $T=0$. A ratio of one indicates
mean-field like superconductivity with negligible  quantum phase
fluctuations, whereas a ratio greater than one shows that quantum
fluctuations are important. In this paper we demonstrate that quantum and thermal fluctuations
also have important consequences for the form of the c-axis conductivity.
In particular, if quantum fluctuations are 
significant, then under conditions believed to occur in
high-$T_c$ materials, $\sigma _{c}(\omega )$ acquires a peak for frequencies 
$\omega >2\Delta $, where $\Delta $ is the superconducting gap. Our previous
spectral weight analysis provided evidence for strong quantum fluctuations
in $T_{c}=70K$ $YBa_{2}Cu_{3}O_{6.6}$ \cite{Ioffe99,Puchkov96} but not in $%
YBa_{2}Cu_{4}O_{8}$ or $YBa_{2}Cu_{3}O_{7}$ \cite{Ioffe99,Puchkov96} and a
peak in $\sigma _{c}(\omega )$ is found in the former material but not in
the latter two \cite{Tajima97}.

The qualitative idea behind our calculations is as follows. We consider
materials, such as the high-$T_c$ superconductors, which consist of weakly
coupled layers. The weak interlayer coupling means that the conductivity may
be calculated by second order perturbation theory in the interplane
coupling, and is therefore given by a convolution of two in-plane Green
functions \cite{Anderson,Chakravarty94}. 
Thus, as emphasized by Anderson \cite{Anderson}, 
the c-axis conductivity is in effect
a spectroscopy of the in-plane properties.  In this paper we show by
explicit calculation in several models what can be learned from
this spectroscopy.
Specific details of the high $T_c$ crystal chemistry imply \cite
{Chakravarty93,Andersen94} that 
in these materials the interlayer coupling is dominated by
the states near $(0,\pi)$ points of the two dimensional Brillouin zone,
where the superconducting gap is maximal.  The c-axis conductivity
thus reflects properties of electronic states near these points.
The in-plane Green function of high $T_c$ superconductors has been studied by
photoemission spectroscopy and in the superconducting state at the momenta
relevant for the interplane conductivity is very small for frequencies
$\omega $ less than the gap energy, $\Delta$ and has a peak
for $\omega \sim \Delta$, (which is interpreted as a
quasiparticle pole), followed by a shallower minimum at an energy $\approx
2\Delta$ followed by a broad continuum, interpreted as the incoherent part
of the spectral function. This structure has been referred to as
``peak-dip-hump''.

In the usual theory of superconductivity \cite{Schreiffer96}, which neglects
phase fluctuations, the type II coherence factors associated with
conductivity mean that the  quasiparticle peak in $\mbox{Im} G(\omega)$
would not contribute to $\sigma_c(\omega)$ which would therefore vanish at
the gap edge, $2 \Delta$, and only begin to rise when the incoherent part of 
$\mbox{Im} G(\omega)$ appears. There would be no sharp structure in $%
\sigma_c(\omega)$ corresponding to the sharp structure in $\mbox{Im}
G(\omega)$. However, we shall show that if phase fluctuations are important,
then the effect of the coherence factors is reduced and the quasiparticles
contribute, leading to a peak in $\sigma_c$ for $\omega \approx 2\Delta$.
The c-axis conductivity is a useful spectroscopy of in-plane properties only
if the tunnelling matrix element is known.  In this paper we assume it has
the usual band theory form, namely an interplane hopping which
conserves in-plane momentum.  A large literature exists in which the
anomalous properties of the c-axis conductivity are attributed to 
a highly non-trivial interplane coupling, in which passage from one plane
to another involves a strong scattering from some excitation or defect
which resides between planes and does not couple to in-plane electron
motion:  see, for example, \cite{Radtke95,Abrikosov96,Kim98}
and references therein.  As discussed in  section VI, we believe
there is substantial
experimental evidence against this proposal.  In any event, the theory
of the effect of superconductivity on $\sigma_c$ in the more straightforward
case of conventional interplane tunnelling and anomalous in-plane properties
is given in this paper.

The rest of this paper is organized as follows. Section II presents a model
for the in-plane Green function introduced by Norman \textit{et al} \cite
{Norman98a} and similar in many respects to a model of Chubukov and
co-workers \cite{Chubukov98}. Section III uses  the model to obtain formulas
for $\sigma _{c}(\omega )$ and shows how quantum and
thermal fluctuations affect the results.
Section IV 
evaluates the results in several limits. Section V discusses the f-sum-rule
spectral weight, section VI  compares our results to data and to alternate
theories, and section
VII is a conclusion and discussion of open problems. Readers uninterested
in the technical details of the calculation are advised to read section II
and then proceed to sections VI and VII.

\section{\textbf{Model:}}

The 'peak-dip-hump' structure discussed above is generally agreed \cite
{Norman98a,Chubukov98,Littlewood92} to imply that the electrons in high-$%
T_{c}$ materials are subject to a strong scattering due ultimately to an
electron-electron interaction -- strong, because the spectral function at
fixed momentum is spread over a wide energy range, and due to
electron-electron interaction because the opening of the superconducting gap
changes the form of the scattering, and in particular weakens it at low
frequencies. Because in optimally doped and underdoped materials
the normal state spectral function is spread over a
wide frequency range and has only a weak structure at fixed momentum near $%
(0,\pi ),$ the imaginary part of the normal state self energy must be large
and only weakly frequency dependent. One theoretical model which leads to
such a self energy involves electrons scattered by some bosonic mode, which 
is thought of as an electronic collective mode and has spectral weight
concentrated near $\omega =0$ in the normal state. If this mode has
electronic origin it must acquire a gap or pseudogap in the superconducting
state. Following refs \cite{Norman98a,Chubukov98} we assume that the mode has
a gap, $\Omega $, at $T < T_c$ where the 'peak-dip-hump' structure is observed,
but has no gap ($\Omega =0$) at $T\geq T_{c}$, where the structure is not observed.
Conventional impurity scattering corresponds to $\Omega=0$ in both normal
and superconducting states. Electron-electron scattering in an s-wave
superconductor at $T=0$ would lead to an $\Omega$ of the order of $2\Delta$
\cite{Littlewood92}
but possibly reduced by excitonic effects. In a two-dimensional d-wave
superconductor (such as high-$T_c$ materials are believed to be), one would
expect a small $\sim \omega^2$ contribution at low frequencies from states
near the gap nodes. We will ignore this small effect, and interpret $\Omega$
as the scale at which the scattering returns to its $T>T_c$ value.

In underdoped cuprates the electron spectral function (and many other properties)
exhibits a 'pseudogap' in a wide range of temperatures above
$T_c$.  The 'pseudogap' is the superconducting energy gap, which seems
\cite{Shen95,Norman98b}
in underdoped materials to persist in a wide
range of temperatures above 
the resistively defined $T_c$.  
In these materials the superconducting transition
corresponds to the onset of long ranged phase coherence
\cite{Emery95,Corson99,Millis99}.  The peak-dip-hump
structure seems to be associated with the establishment of phase coherence,
and not with the formation of the gap.  This behavior 
is not at present understood
\cite{Millis99}.  In the present paper we simply assume 
it occurs and examine its
consequences for the c-axis conductivity.

For several reasons, including that discussed just above,
the properties of the mode required by the models
of Refs \cite{Norman98a,Chubukov98} seem somewhat unusual,
raising the question of whether a description in terms of scattering
of conventional electrons off of a bosonic mode is the physically
correct one.  The models however are reasonably successful in fitting
the electron spectral function, and this is all that we require here.
The issue of the proper physical picture of the unusual
behavior of the electron spectral function is, however, a crucial
issue in the physics of high temperature superconductivity.  We return to 
it in the conclusion. We now turn to the mathematical formalism we need.

We may write in general for the normal ($G$) and anomalous ($F$) propagators
in Matsubara formalism:

\begin{equation}
G(p,i\omega_n )=\frac{-i\omega _{n}Z_{p}(\omega _{n})-\epsilon _{p}}{(\omega
_{n}^{2}+\Delta_{\omega_n }(p)^{2}))Z_{p}(\omega _{n})^{2}+\epsilon _{p}^{2}}
\label{G}
\end{equation}

and

\begin{equation}
F(p,i\omega_n) = \frac{\Delta_{\omega_n}(p)Z_{p}(\omega _{n})}
{(\omega_n^2+\Delta^2_{\omega_n}(p))Z_p(\omega_n)^2+
\epsilon_p ^2}  \label{F}
\end{equation}

Here $G(p,\omega_n)=\int d\tau e^{i \omega_n\tau } \langle T_{\tau} c_p^{\dagger} 
(0) \rangle $,  $F(p,\omega_n)= \int d\tau e^{i \omega_n\tau } 
\langle T_{\tau} c_p (0) \rangle $ and 
$Z_p(\omega )$ is the renormalization function defined by $i\omega -\Sigma
(p,i\omega )=i\omega Z_p(i\omega )$ where $\Sigma (i\omega )$ is the self
energy which contains the effects of coupling to the mode which produces the
strong normal-state scattering. The approximation of 
Norman et. al. \cite{Norman98a}, which is adequate for 
our purposes and which we adopt henceforth, 
consists of neglecting the
frequency dependence of $\Delta $ and the momentum dependence of $Z$. The
frequency dependence of $\Sigma $ is given by

\begin{eqnarray}
\Sigma (i\omega_n )
&=&\frac{\Gamma }{2\pi }\frac{\omega_n -i\Omega }{\sqrt{(\omega_n
-i\Omega )^{2}+\Delta(p) ^{2}}} \nonumber  \\ &\mbox{ln} &
\left[ \frac{\omega_n -i\Omega +\sqrt{(\omega_n
-i\Omega )^{2}+\Delta(p) ^{2}}}{i\Delta(p) }\right] -cc  \label{Z}
\end{eqnarray}
This form corresponds to a single-particle scattering rate which tends
to $\Gamma/2 $ for $\omega \gg \Omega ,\Delta $
and reduces to the familiar expressions for a dirty superconductor
when $\Omega \rightarrow 0$. The observed \cite{Homes95}
frequency independence of $\sigma _{1}$ at $T>T_{c}$ 
in the range $\omega <1/2eV$ leads us to choose 
$\Gamma \sim 0.8eV \gg \Delta $. 
At large frequencies
we therefore have very strongly scattered electrons, 
corresponding to an $ImG(p,\omega)$ which is small $\sim 1/\Gamma $ 
and essentially independent of $p,\omega $.  
However, for $\omega < \Delta +\Omega$
Z is real, so 
G will have a pole at a frequency 
$\omega_{qp}=\sqrt{\Delta^2+\epsilon_p^2/Z(\omega_{qp})^2}$
if $\omega_{qp} <\Delta+\Omega$.  A large $\Gamma$ such
as we have assumed
implies  that $Z(\omega_{qp} \gg 1)$ so the quasiparticles
have small weight and negligible dispersion. Refs 
\cite{Norman98a,Chubukov98,Norman98b} argue that 
the combination of a large $\Gamma$ and an $\Omega=0$
in the superconducting state but not in the normal state
accounts  for the peaks observed in 
photoemission experiments at $T<T_{c}$
but not for $T>T_{c}$ \cite{Shen95,Norman98b}.
  
  The resulting spectral function is shown in Fig.~1 for $\epsilon_p=\Delta$
and $\Gamma=40 \Delta$. Here the quasiparticle pole is 
shown as a sharp line and its strength corresponds to the area in
the shaded box.  The onset of scattering at $\omega=\Omega+\Delta$
causes the broad only weakly frequency dependent 
continuum.

\vspace{0.25cm}
\centerline{\epsfxsize=9cm \epsfbox{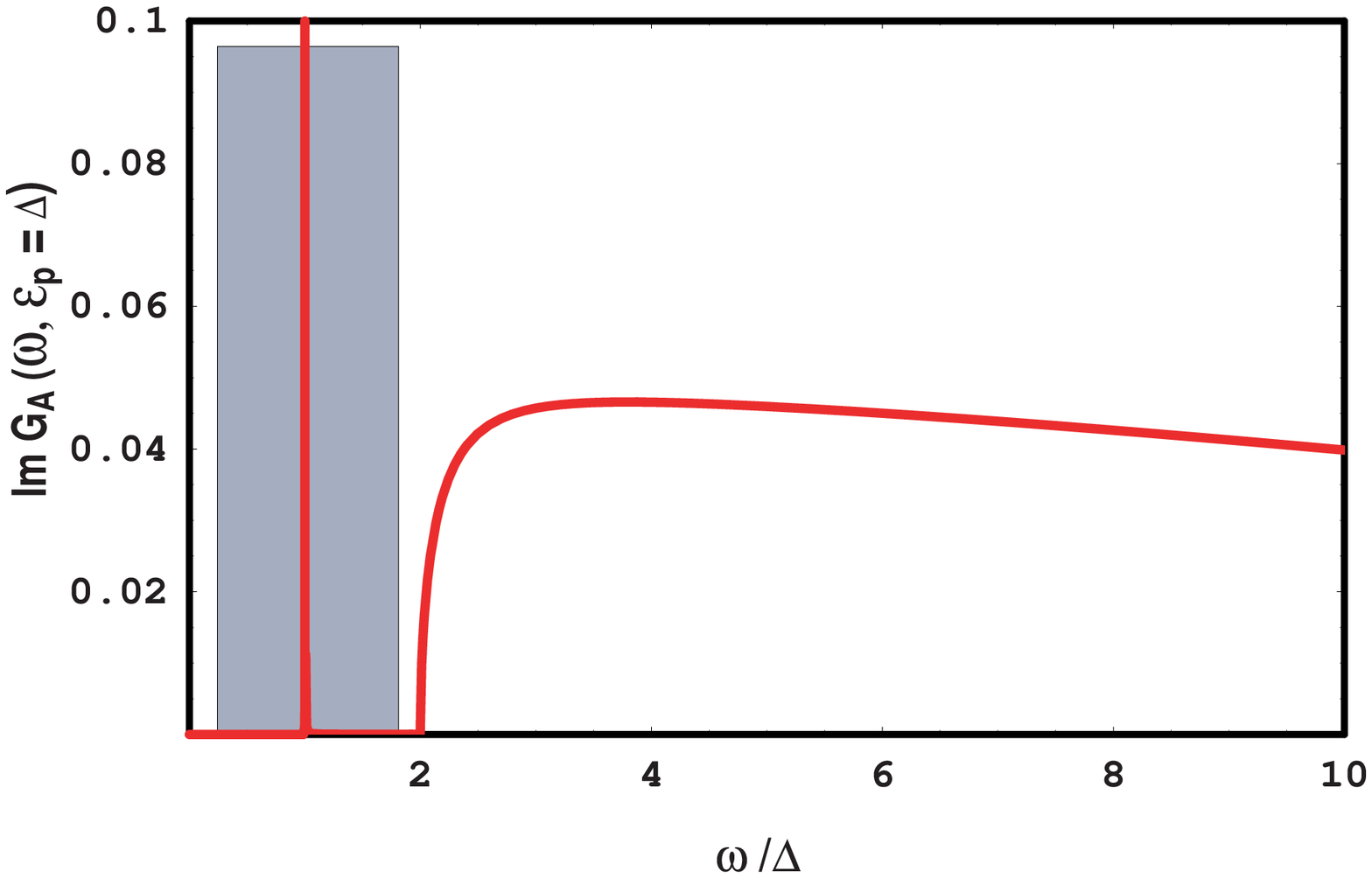}} 
{\footnotesize \textbf{Fig. 1} Imaginary part of the electron Green function
for parameters $\Gamma=40$ and $\Omega=\Delta$.  The quasiparticle peak is
indicated by the vertical line; its weight is shown as the 
shaded box.} 
\vspace{0.25cm}

The spectral function predicted by Eq. \ref{Z} was fit to photoemission data
by Ref \cite{Norman98a}. Differences between the data and 
the model were attributed by 
Ref \cite{Norman98a}
to an extra elastic scattering term and to an extrinsic background. The
elastic scattering was introduced to broaden the low frequency peak,
which is a delta function in the model; we feel the 
observed broadness is more likely due to a variation 
of the gap over the surface
of the sample, but in any event this extra broadening (which we do not
include) will not affect our results in any significant way.

The extrinsic background requires more discussion. 
In the data analysed in Ref \cite{Norman98a}
the weight in the quasiparticle peak was somewhat
smaller than the
'missing' area obtained by multiplying the higher 
$\omega$ value of Im G  by
the frequency interval $0 < \omega <\Delta+\Omega$.  
As can be seen from our
Fig. 1, in the strong coupling limit of the model self 
energy the strength of
the quasiparticle peak is in fact somewhat larger than 
this missing area.
The authors of Ref. 7 apparently dealt with this discrepancy 
by introducing an
additional extrinsic background, non-zero only for 
$\omega > \Delta+\Omega$,
and adjusted so the weight in the 
observed quasiparticle peak is 
roughly equal to the 'missing'
area calculated from the background-subtracted part of Im G. 
It seems to us that the
discrepancy requires further consideration; however, the 
issue is not crucial
to the present paper, which focusses on the qualitative 
consequences 
for the c-axis conductivity of the
quasiparticle peak in Im G.

\section{\protect\smallskip Interplane Conductivity}

We now consider the effect of our chosen  form of $G$ and $F$ on 
$\sigma _{c}$, focussing on the extent to 
which the quasiparticle poles contribute to
the observed conductivity and on the effects of the offset $\Omega$. 
We emphasize that for these considerations the
precise forms of $G$ and $F$ do not matter, as long as they have the general
properties outlined above. We assume the Hamiltonian is
\begin{equation}
H=H_{in-plane}+\sum_{p,\sigma,i}(t_{\perp}(p)c_{p,\sigma,i}^{\dagger}
c_{p,\sigma,i+1} +H.c.)
\label{H}
\end{equation}
Here $c_{p,\sigma,i}^{\dagger}$ creates an electron of in-plane momentum
$p$ and spin $\sigma$ on plane $i$.  $H_{in-plane}$ contains the (presumably
nontrivial) physics of a single $Cu-O_2$ plane, and in particular
leads to the $G$ and $F$ functions discussed in the previous section.
We calculate the interplane conductivity in the usual way, representing
the c-direction electric field by a vector potential $A$ and
coupling it to $H$ via the Peierls substitution (in units 
$\hbar = c = 1$)
$t_{\perp} \rightarrow t_{\perp}e^{i e A d}$ with d the interplane spacing.
We expand to second order in $t_{\perp}$ finding 
\cite{Ioffe99,Anderson,Chakravarty94}

\begin{equation}
\sigma _{c}(\omega )=\frac{1}{i\omega }
(K-T\sum_{\omega^{'} }\int d\widehat{p}%
t_{\perp }^{2}(\widehat{p})\Pi (\omega^{'} +
\frac{\omega }{2},\omega^{'} -\frac{\omega }{2},\widehat{p}) )
\label{sigma_c}
\end{equation}

We have omitted dimensional factors of $e^2$ and lattice constants,
which are not relevant to our arguments.
It is convenient to consider separately contributions to 
the polarizibility $\Pi $ coming from the different parts of 
the Fermi surface (as labeled by $\widehat{p}$) 
and separate $\Pi $ into normal and anomalous parts, as 

\begin{equation}
\Pi=\Pi_{GG}-\Pi_{FF}
\label{Pidfn}
\end{equation}
with 
\begin{eqnarray}
\Pi _{GG}(\omega^{'} _{+},\omega^{'} _{-}) &=&
\nu \int d\xi G_{i}(p,\omega^{'}
_{+})G_{i+1}(p,\omega^{'} _{-}) \\
\Pi _{FF}(\omega^{'} _{+},\omega^{'} _{-}) &=&\nu \int d\xi 
\langle F_{i}(p,\omega^{'}
_{+})F_{i+1}(p,\omega^{'} _{-})\rangle   
\label{Pi}
\end{eqnarray}

Here $\xi=v_F(p-p_F)$, $\omega^{'}_{\pm}=\omega^{'} \pm \omega$
and here and in many places subsequently
the $\widehat{p}$ label on the $\Pi$ functions has
been suppressed. 
The diamagnetic term K may be written
\begin{equation}
K=T\sum_{\omega }\int d\widehat{p}
t_{\perp }^{2}(\widehat{p})\Pi_{GG} (\omega  ,\omega ,\widehat{p}) )+
\Pi_{FF} (\omega  ,\omega ,\widehat{p}) )
\label{K}
\end{equation}

We have written the FF term as an expectation value because
it depends on the interlayer phase coherence.To study the phase 
coherence properties in more detail we write the F-F
correlator in real space and separate out the term involving the phase
difference, finding
\begin{equation}
\Pi_{FF}(r,t)=\Pi_{FF}^0(r,t)
\langle e^{i\phi_i(r,t)-i\phi_{i+1}(0,0)}\rangle
\label{PiFF}
\end{equation}
Here $\Pi^0_{FF}$ is the usual convolution of 
F-functions with phases set to
0 and may be calculated by standard methods \cite{Schreiffer96}.
In the BCS approximation the phases do not 
fluctuate: $\phi _{i}(r,t)=\phi_{i+1}(r^{\prime },t^{\prime })=\phi _{0}$
and $\Pi_{FF}=\Pi_{FF}^{0}$. In
the actual materials  the phase in each plane fluctuates.  
In the 'pseudogap'
regime of underdoped materials the superconductivity is destroyed 
by phase fluctuations while the amplitude of the gap remains 
nonzero \cite{Emery95,Corson99,Millis99}.
For $T <T_c$ the phase has a nonzero average but may fluctuate.  
We write $\phi (r,t)=\phi _{0}+\delta \phi (r,t)$. At $T<<T_{c}$
thermal fluctuations are negligible. In the two-dimensional case of present
interest, quantum fluctuations are dominated by short length scales and
so are uncorrelated from plane to plane, $\langle \delta \phi _{i}\delta
\phi _{i+1}\rangle =0$. This allows us to evaluate the correlator entirely
in terms of in-plane properties, at low $T$. Unlike quantum fluctuations,
two-dimensional thermal fluctuations are dominated by length scales of the
order of the thermal coherence length $\xi _{T}$ which diverges at $T_{c}$.
Interplane phase correlations are important if the Josephson energy of a
correlated region is larger than the temperature, 
i.e. if $N_{0}t_{\perp}^{2}\xi _{T}^{2} \Delta /\Gamma>k_{B}T$; 
where $N_{0}$ is the in-plane density of states and the 
standard dirty-limit factor $\Delta/\Gamma \ll 1$ 
reflects the fact that the weight in the c-axis 
conductivity is spread
over a wide frequency range of order $\Gamma$, so only 
the fraction $\Delta/\Gamma$ is available to 
contribute to the Josephson coupling.
The small values of interplane couplings 
and $\Delta/\Gamma$ relevant to high-$T_{c}$ materials
mean that this criterion is only satisfied for 
temperatures very near $T_{c}$. Thus except very near to $T_{c}$ we have
\begin{equation}
\Pi _{FF}(r,t)=\Pi_{FF}^{0}(r,t)
e^{-\langle (\delta \phi _{i})^{2}\rangle}
\equiv \alpha \Pi_{FF}^{0}(r,t)  
\label{alpha}
\end{equation}
In other words in $d=2$
the effect of quantum and thermal phase fluctuations is to
renormalize the interplane $F-F$ correlator by a 
constant Debye-Waller factor $\alpha$ 
with $0<\alpha <1$.  In $d=1$ the phase fluctuation integral has a logarithmic 
divergence cut off by $t_{\perp}$ or the length and time scale, so the
Debye-Waller factor will have scale or $t_{\perp}$ dependence.

We shall be interested in either a strongly scattered normal state
or in $T \ll \Delta$ so we take the $T \rightarrow 0$
limits of the formulas.
After integration over momentum and analytical continuation we find for the
imaginary parts

\begin{eqnarray}
Im \Pi_{FF}^0(\omega^{'}_{+},\omega^{'}_{-}) &=& \frac{1}{2} 
Re\left[ \frac{\Delta^2}{(\xi_{+}^R Z_{+}^R + \xi_{-}^R Z_{-}^R) 
(\xi_{+}^R \xi_{-}^R)} \right. 
\nonumber \\
&-& \left. \frac{\Delta^2}{(\xi_{+}^R Z_{+}^R + \xi_{-}^A Z_{-}^A)
(\xi_{+}^R \xi_{-}^A)} \right]  \label{Pi_FF}
\end{eqnarray}
and
\begin{eqnarray}
Im \Pi_{GG}(\omega^{'}_{+},\omega^{'}_{-}) &=& \frac{1}{2} Re\left[ \frac{1}{%
\xi_{+}^R Z_{+}^R + \xi_{-}^R Z_{-}^R} \left(1+\frac{\omega_{+}\omega_{-}}{%
\xi_{+}^R \xi_{-}^R} \right) \right.  \nonumber \\
&-& \left. \frac{1}{\xi_{+}^R Z_{+}^R + \xi_{-}^A Z_{-}^A} \left(1+\frac{%
\omega_{+}\omega_{-}}{\xi_{+}^R \xi_{-}^A} \right) \right]  \label{Pi_GG}
\end{eqnarray}
where $\xi_{\pm}^R = \sqrt{\Delta^2 - (\omega^{'}_{\pm}+i \delta)^2}$, $%
\xi_{\pm}^A = (\xi_{\pm}^R)^*$  
and $Z_{\pm}^R =1-\Sigma^R(\omega^{'}_{\pm})/\omega^{'}_{\pm}$.

Finally, to compute $\sigma_c$ we must substitute 
Eqs \ref{Pi_FF}, \ref{Pi_GG}
into Eq. \ref{sigma_c} and average over the Fermi surface.  
As mentioned above,
in high-$T_c$ materials $t_{\perp}$ is very strongly peaked about 
the $(\pi,0)$
points where the superconducting gap is maximal. Thus we may approximate
$\Delta(p)$ by its maximum value $\Delta$ and ignore the d-wave gap structure
and the integral over angles. We then obtain  for
the absorptive part of the conductivity
\begin{equation}
\sigma^{(1)}(\omega)=\frac{\sigma_0}{\omega}
\int_{-\frac{\omega}{2}}^{\frac{\omega}{2}} \! d\omega^{'} 
\mbox{Im} [\Pi_{GG} (\omega^{'}_+,\omega^{'}_-)
-\alpha \Pi_{FF}(\omega^{'}_+,\omega^{'}_-)]
\label{sigma1}
\end{equation}
with $\sigma_0=N_0\int d\widehat{p} t_{\perp }^{2}(\widehat{p})$.

\section{Evaluation of conductivity}

  We begin with the case $\Omega=0$ i.e. with strong scattering 
unaffected by the onset of phase coherence. Results
are shown in Fig. 2
For this choice of $\Omega$,
the real part of the normal state conductivity is
$\sigma_c(\omega)=\Gamma/ (\omega^2 +\Gamma^2)$. 
We have chosen a very
large $\Gamma$ so the normal conductivity is a straight line with value 
$\Gamma \sigma=1$.  The light solid line depicts the conductivity of the
fully phase coherent ($\alpha =1$)
superconducting state.  The coherence factor effects,
namely $\sigma(\omega = 2\Delta)=0$ and the
gradual onset of absorbtion 
as $\omega$ is increased above $2 \Delta$, are evident.  
Also shown as the heavy
solid line in Fig. 2 is
the no-phase coherence ($\alpha=0$) conductivity .  
The coherence factor effects are absent, so the conductivity 
rises discontinuously from the gap edge (in our approximation,
which neglects angular variations of the gap) and is always larger than the
phase-coherent $\sigma$.  
No sharp structure is visible in either calculation 
because everywhere an excitation is allowed, the damping is large.

In the $\Omega=0$ case the conductivity in the presence of the gap
is always less then the normal state ($\Delta=0$) conductivity.
In the fully phase coherent ($\alpha=1$) case 
there is also a delta function contribution (not shown); the
weight in the delta function exactly equals the difference of the fully phase
coherent curve from 1. 
In the $\alpha=0$ case the lack of phase coherence means that there is no
superfluid delta-function--the conductivity spectral weight is less than in
either the no-gap case or the fully phase coherent state. The spectral weight
will be discussed in more detail in the next section. 

Because it will be useful in our subsequent discussions,
we also show as the dashed line in Fig. 2
the conductivity calculated assuming no
scattering ($\Gamma=0$) and no phase coherence ($\alpha=0$), but with a 
non-vanishing superconducting gap. (This conductivity is
normalized in a way which does not involve $\Gamma$). 
In this case quasiparticle absorbtion
above the gap edge is allowed, as in a usual semiconductor, 
even though there is no scattering, leading to the square root divergence
shown.

\vspace{0.25cm} 

\centerline{\epsfxsize=9cm \epsfbox{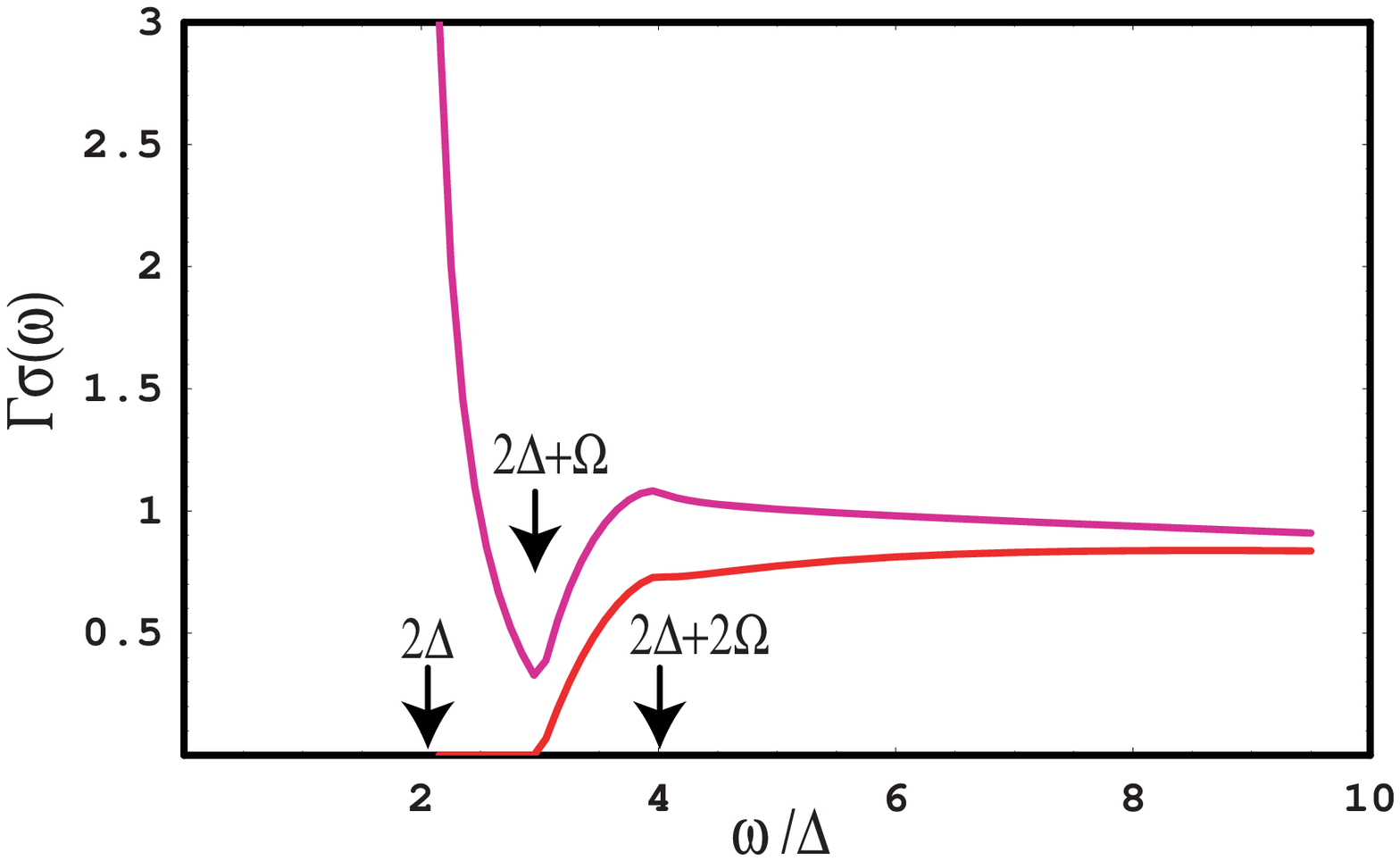}} 

{\footnotesize \textbf{Fig. 2} Optical conductivity multiplied
by scattering rate $\Gamma$ with 
(thin dark line) and without (thick light line) phase
coherence calculated for $\Omega=0$, 
in the limit $\Gamma \rightarrow \infty$.
The normal state (no gap) conductivity $\Gamma \sigma =1$. Also shown
as dashed line is conductivity from un-scattered quasiparticle
states in absence of phase coherence.}
 \vspace{0.25cm}
 
We now turn to the effects of the offset. We begin by considering
in Fig. 3 the conductivity of a non-paired 
($\Delta=0$) state at $T=0$, for different values of $\Omega$.  
To facilitate
comparison with subsequent figures we measure frequencies and 
the offset in
units of $\Delta$, but we emphasize 
that these curves pertain to the non-paired 
state.  At $\Omega=0$ we would have a Lorentzian of half-width $\Gamma=40$
and value 1 at $\omega=0$;
this is not shown in the Figure. We see from the three displayed curves
that $\sigma(\omega < \Omega)=0$ (except for a delta-function contribution
of relative strength $\Gamma/Z(\omega=0)=\frac{1}{1+\Gamma/\pi \Omega}$)
which is not explicitly shown). The conductivity has a complicated dependence
on the combination of scattering rate and offset:  if $\Omega/\Gamma$ is
sufficiently small (as occurs in the displayed $\Omega=0.5\Delta$
and $\Omega=\Delta$ traces)
the conductivity can rise above the no-offset value (basically, $\sigma=1$) 
for a $\Gamma$-dependent range of $\omega$; for larger offsets (e.g
the $\Omega=2\Delta$) trace, the conductivity is always lower.  The total
weight (including the delta-function contribution) 
is conserved as a function of $\Omega$, but as can be seen from the
high frequency behavior of the curves, the differences between
$\sigma$ calculated with different $\Omega$ persist 
up to very high frequencies,
so making accurate statements about differences in integrated area
requires integration over frequencies of order $\Gamma$.

\vspace{0.25cm} 
\centerline{\epsfxsize=9cm \epsfbox{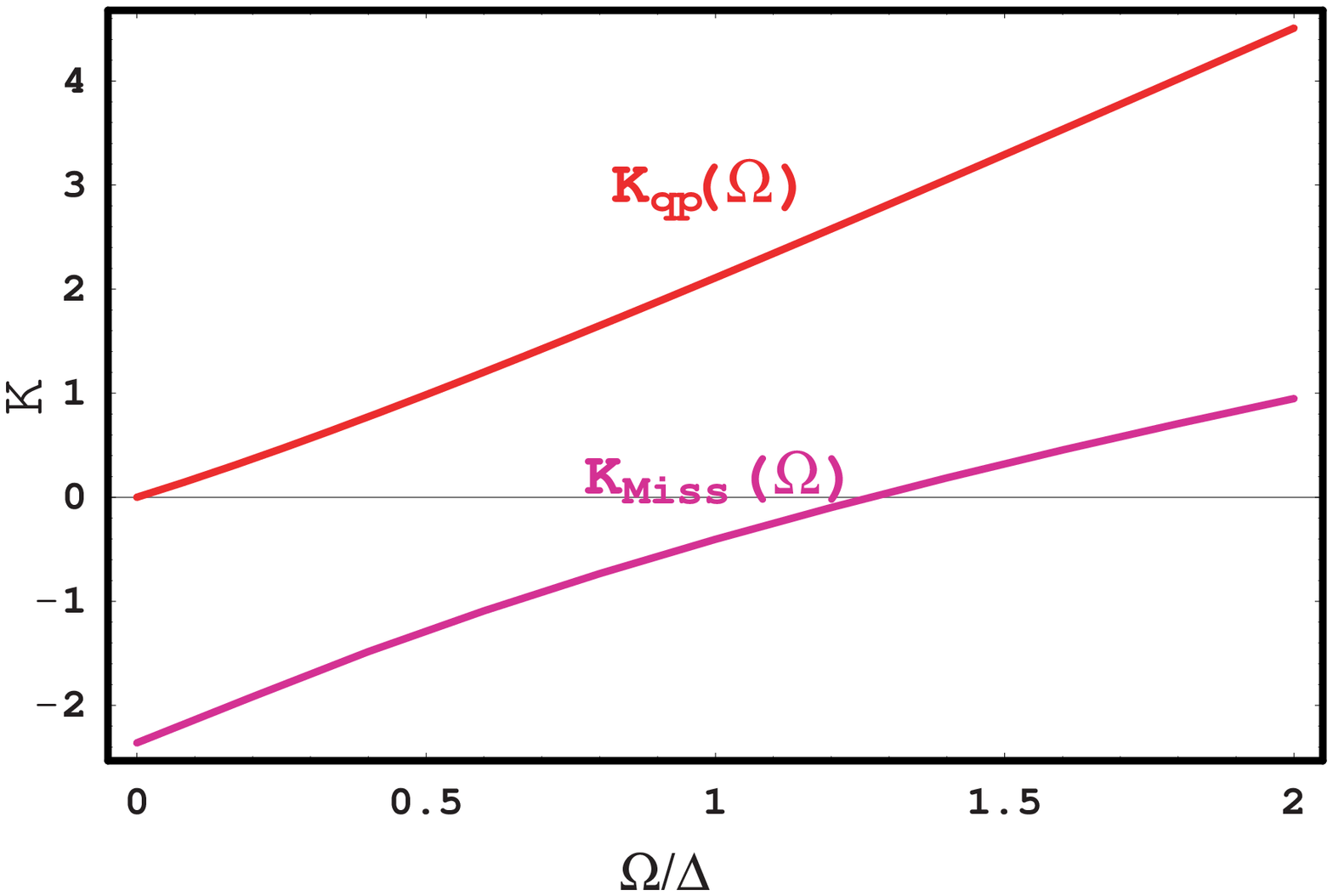}} 
{\footnotesize \textbf{Fig. 3} Optical conductivity in normal state
as function of mode offset $\Omega$
for $\Gamma=40$. }
\vspace{0.25cm}

We next show in Fig. 4 the conductivity in the superconducting state,
with offset $\Omega=\Delta$ and $\Gamma=40$. 
In this case in the energy range
$\Delta < \omega < \Delta +\Omega$ unscattered quasiparticles exist.
In the case of
perfect phase coherence ($\alpha=1$) we obtain the lower curve.
The onset of the absoption at $\omega = 2 \Delta + \Omega$
is evident, as is the restoration of the full scattering at $\omega > 2
\Delta + 2 \Omega$. The freely propagating quasiparticle states at $%
|\omega|<\Delta+\Omega$ do not contribute to $\sigma$ because of the type II
conductivity coherence factors. In physical terms, these states are not
scattered and one gets no absorption without scattering. Comparison
to the appropriate curve in Fig. 3 shows that convergence to the
$\Delta=0$ case of the same $\Omega$ is very slow.  We have verified that
the area including the superfluid delta function (not shown) is conserved,
but one must integrate to frequencies of order $\Gamma$ to 
capture all of the spectral weight.

We now consider the opposite case, namely that the interplane quantum
fluctuations are so strong that $\Pi_{FF}$ is negligible. In this case $%
\sigma_c$ is determined from $\Pi_{GG}$ alone and the resulting conductivity
is shown as the upper line in Fig.~4. A quasiparticle contribution is
evident in the region $2\Delta+\Omega > \omega >2\Delta$; this has relative
weight $1/Z \ll 1$ and in the simple approximation considered here has
the same functional form (and physical origin) as the dashed curve shown
in Fig. 3.. The threshold of inelastic
scattering at $\omega=2 \Delta+\Omega$ is evident as is the restoration of
the full scattering at $\omega=2 \Delta + \Omega$. The absorbtion at $2
\Delta < \omega < 2 \Delta + \Omega$ is at first sight surprising because it
comes from freely propagating quasiparticles which normally cannot lead to
finite frequency absorbtion in a translationally invariant system. Such
absorbtion can be easily understood if the interplane phase coherence is
destroyed by the thermal fluctuations. In this case we may think of each
plane as having a separate phase, so that translation invariance is
effectively broken. In the $T=0$ case of present interest, however, one
deals with the ground state of a quantum system which has a translation
invariance. The explanation in this case is that $T=0$ interplane
fluctuations imply the existence of an interplane charging energy; the
presence of this term in the Hamiltonian means that the current carried by
the quasiparticles does not commute with the Hamiltonian and this allows the 
$\omega \sim 2 \Delta$ absorption to exist.
\vspace{0.25cm} 
\centerline{\epsfxsize=9cm 
\epsfbox{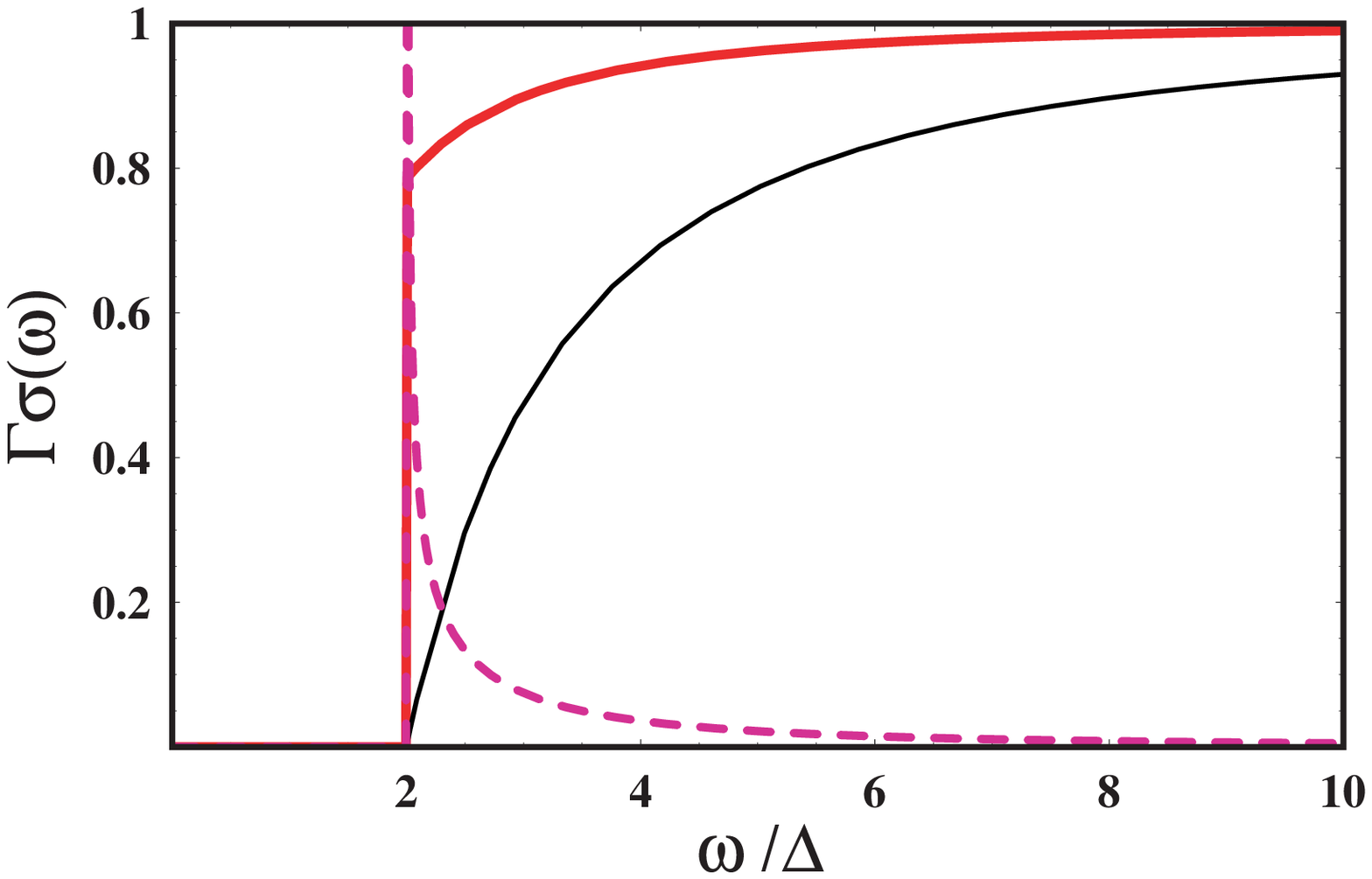}} 

{\footnotesize \textbf{Fig. 4} 
Optical conductivity in superconducting state with (lower line) and without (upper line) 
phase coherence, for $\Gamma=40$ and $\Omega=\Delta$.}
\vspace{0.25cm}

\section{\protect\smallskip Spectral Weight}

In this section we study in more detail   the f-sum rule spectral weight,
i.e. the integrated area under the conductivity.
Standard analysis
\cite{Ioffe99,Maldague77,Baerswyl87,Millis90} 
shows that this is related to K, Eq. \ref{K}, via
$K=\sigma_0 \int \frac{2d\omega}{\pi} \sigma(\omega)$. The results of
section IV imply
\begin{equation}
K(\Omega,\Delta,\alpha)=\sigma_0\int^{\infty}_0 \frac{2d\omega}{\pi}
\int d\xi_p [G(p,\omega)^2+\alpha |F(p,\omega)|^2]
\label{Kexplicit}
\end{equation}
In the superconducting state, some of the weight is concentrated in
a delta-function at $\omega =0$.  The coefficient of this delta function
is conventionally written as $\pi \rho_s$ and is given by the difference
between $K$ and $\Pi(\omega \rightarrow 0)$ \cite{Schreiffer96}.  We find
\begin{equation}
\rho_s(\Omega,\Delta,\alpha)=2 \alpha \sigma_0
\int^{\infty}_0 \frac{2d\omega}{\pi}
\int d\xi_p |F(p,\omega)|^2
\label{rhosexplicit}
\end{equation}

These results depend crucially on the assumption that the only
important interplane hopping term is that written in Eq. \ref{H}
(arbitrary interplane interactions are allowed).  Different
hopping terms, such as those considered in 
\cite{Radtke95,Abrikosov96,Kim98,Hirsch92} would lead to different results,
as noted in \cite{Kim98,Hirsch92}.

It is evident from Eqs \ref{Kexplicit},\ref{rhosexplicit} that if
all other parameters remain fixed, then a decrease in $\alpha$ causes
related decreases in the total spectral weight and $\rho_s$.  It is natural
to assume that if $\alpha=1$ (no phase fluctuations) then spectral
weight is conserved as a function of temperature, so that variations
in spectral weight as a function of temperature imply variations
in $\alpha$.  In a
previous paper \cite{Ioffe99} we studied the consequences of
this assumption.
We distinguished three cases: (i) no pairing(i.e. $%
\Delta =0$ and $\sigma \sim \int GG$); (ii) conventional (no fluctuations)
superconductivity (i.e. $\Delta \neq 0$ and $\sigma \sim \int
GG-FF$); (iii) superconductivity with strong phase fluctuations (i.e. $%
\Delta =\neq 0$ and $\sigma \sim \int GG-\alpha FF$, $\alpha \ll 1
$). Cases (i) and (ii) were found to have the same total spectral weight but
going from case (ii) to case (iii) by keeping the gap the same but
increasing the phase fluctuations (i.e. decreasing
$\alpha$) was shown to decrease the total spectral
weight. A particularly interesting case was $\alpha =0$ which has been
argued to represent the $T>T_{c}$ pseudogap regime of underdoped cuprates 
\cite{Emery95,Ioffe99}. Reducing $T$ from room temperature,
where no pseudogap is evident, to $100-150K$, where a pseudogap is plainly
seen in many measurements in underdoped cuprates, corresponds to going from
case (i) to case (iii), i.e. it reduces the total spectral weight. Further
reducing $T$ to a $T<T_{c}$ induces long-ranged phase coherence, implying $%
\alpha >0$ and hence an increase in spectral weight. If as $T\rightarrow 0$
phase coherence is fully restored at all scales, i.e. if quantum
fluctuations are negligible, we may set $\alpha =1$. In this case the $T=0$
spectral weight equals the high-temperature spectral weight and in
particular $\rho _{s}$ is given by the area lost in $\sigma _{c}(\omega >0)$
between high temperature and $T=0$. A $T=0$ value of $\rho_s$ which is less
than the "missing area"  was therefore argued \cite{Ioffe99}
to imply non-negligible quantum
fluctuations.

The results of \cite{Ioffe99} relied on the 
$T$-independence of the spectral weight at $\alpha=1$.  
This was verified in a
BCS like model \cite{Ioffe99} but the complicated interplay between
superconductivity and scattering indicated in 
\cite{Norman98a,Chubukov98,Littlewood92} means that additional analysis is
required.  We show here that the crucial assumption is that the self-energy
function $Z_p(\Omega)$ has negligible  dependence 
on band energy $\xi_p$ in the frequency
ranges of interest. We consider the difference between
$K(\Omega,\Delta,\alpha)$ and the noninteracting spectral weight 
$K_{nonint}$, which is given
by Eq. \ref{K} with $Z=1$.  
Because the
high frequency, large $\xi$ asymptotics of the two integrands are the same,
we may integrate over $\xi_p$ first, using the $\xi_p$-independence
of $Z$ and obtaining
\begin{equation}
K(\Omega,\Delta,\alpha)-K_{nonint}=\sigma_0 \int^{\infty}_0 \frac{2d\omega}{\pi}
\frac{\pi (\alpha-1) \Delta^2}{2 Z (\omega^2+\Delta^2)^{3/2}}
\label{delK}
\end{equation}

When $\alpha=1$ the right hand side of Eq. \ref{delK}
vanishes: spectral weight is conserved as a function of $\Omega$ and $\Delta$. 
We note that because this integral is dominated by $\omega \sim \Delta$
the crucial requirement is that $Z_p(\omega)$ have negligible
$\xi_p$ dependence for $\omega \sim \Delta$.

Using the same assumption we find for the superfluid stiffness

\begin{equation}
\rho_s=2 \alpha \sigma_0 \int^{\infty}_0 \frac{2d\omega}{\pi} \frac{\Delta^2}
{Z (\omega^2+\Delta^2)^{3/2}}
\label{rhos}
\end{equation}

The $\Omega$ dependence of $\rho_s$ is shown in Fig. 5 for $\Gamma=40$
and $\alpha=1$. One sees that $\rho_s$ increases rapidly with increasing 
$\Omega$. The physics may be understood most
simply from the $\Omega$-dependence of $\rho_s$. For impurity scattering 
($\Omega=0$) $\rho_s$ is reduced from the non-interacting value by the usual
dirty limit factor $\Delta/\Gamma$; this is properly understood as a mass
renormalization $\rho_s \to \rho_s/Z$ coming from a self energy which (for $%
\omega < \Delta$) is real but has a strong frequency dependence ($\Sigma
\approx \omega Z \approx \omega \Gamma/ \Delta$). If we now add a threshold $%
\Omega$ to the scattering mechanism the renormalization decreases: $Z
\rightarrow \Gamma/(\Delta+\Omega)$; the 
decrease in $Z$ leads to an increase in $\rho_s$.

\vspace{0.25cm} 
\centerline{\epsfxsize=9cm 
\epsfbox{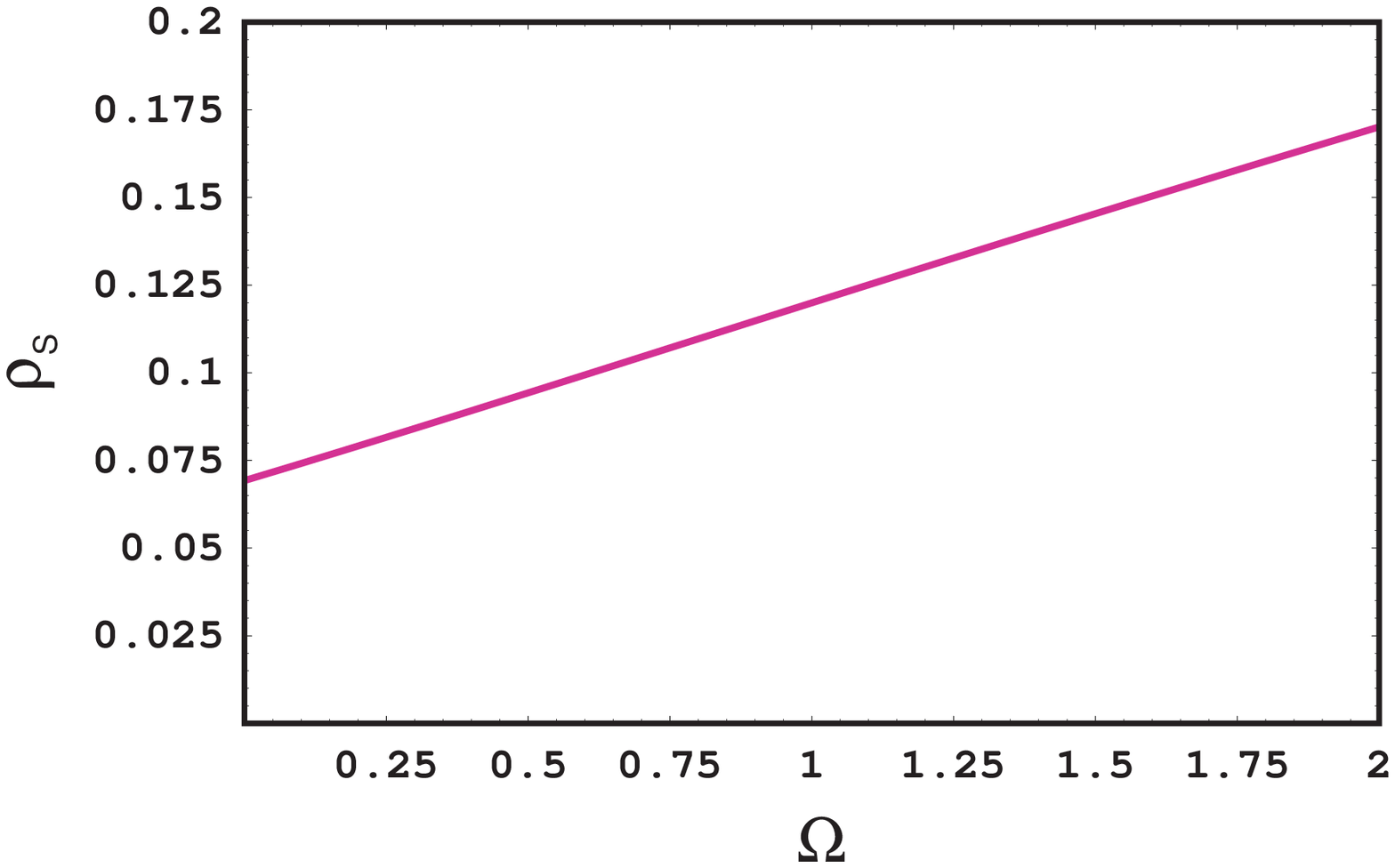}} 

{\footnotesize \textbf{Fig. 5} Dependence of superfluid stiffness on
offset $\Omega$ for $\Gamma=40$.  
These numbers must be multiplied by $\Gamma/\pi$
in order to be compared directly to areas from scaled conductivities
shown in previous figures.}

A similar $\omega$ dependence occurs in the spectral weight of the
``quasiparticle'' peak in the $\alpha=0$ conductivity, again because
the quasiparticle mass $\sim 1/Z$.
For example, in the case $\Omega=\Delta$  the weight in the 
$\alpha=0$ quasiparticle
peak is approximately equal to the weight in the conductivity at high $T$
(i.e. $\Delta=0$) over the range $(0,2\Delta)$. Of course, this is less
than the total 'missing weight'(indeed is about half) 
because $\sigma$ is suppressed
over the wider range $\omega \sim 2\Delta+\Omega$.

For $\alpha=1$,weight is conserved as $\Omega$ is varied, so the 
increase in $\rho_s$ is compensated by a decrease in the $\omega >0$ 
conductivity. We see however from the normal state calculation shown in 
Fig 2 that as $\Omega$ is changed the difference in weight
is spread out over a wide range of frequencies, of order $\Gamma$.  If one
integrates $\sigma$ over a range small compared to $\Gamma$ the
total weight  ($\rho_s$ plus $\omega >0$ part) may 
appear to increase as $\Omega$ is increased.  By contrast, we find that
if $\Delta$ is varied at fixed $\Omega$. the change in conductivity
is more concentrated at frequencies of the order of $\Delta$ and
weight is to a reasonable approximation conserved even if one integrates
only over a range of the order of a few $\Delta$, especially
for $\alpha < 1$.

Weight is not conserved if the Debye-Waller parameter $\alpha$ is varied.
Varying $\alpha$ changes the weight in the superfluid component, 
the weight in the
non-zero-frequency component, and also  the total weight.
In our previous work \cite{Ioffe99}
we found (as can be seen directly from Eqs. \ref{delK},\ref{rhos})
 that increasing $\alpha$ while keeping 
other parameters fixed increases the weight in the c-axis 
superfluid response by
twice as much as the total spectral weight increase.
The physics is that as $\alpha$ 
is increased, coherence factor effects become more important, leading
to a decrease in $\sigma_c(\omega)$ in the region $\omega>2 \Delta$; this
decrease reduces the non-zero frequency spectral weight 
$\int_{\omega=0^{+}} \sigma(\omega)
d \omega$ by an amount which turns out to be 
half of the increase in $\rho_s$.
We are most intersted in applying this result to underdoped cuprates, 
in which the superconducting gap is well formed for temperatures
of the order of the resistive $T_c$ and the onset of superconductivity
leads to an increase in $\alpha$ from 0 to some non-zero
value. However, as we have seen, there is evidence that the onset of
superconductivity leads also to a change in $\Omega$;
therefore further consideration is necessary.

Further consideration requires further assumptions.  We focus on
underdoped materials and assume that at temperatures slightly greater
than $T_c$ the superconducting gap is well formed and much greater than
$T$, but that thermal phase fluctuations drive $\alpha$ to $0$.
We further
suppose that at $T>T_c$ the offset parameter $\Omega=\Omega_+$.
As $T \rightarrow 0$ long ranged phase coherence is established, so
$0 < \alpha \leq 1$ and $\Omega \rightarrow \Omega_-$. 
The data seem consistent with $\Omega_+=0$ and
$\Omega_- \approx \Delta$ but we prefer to present
a more general treatment.  
We now use Eqs. \ref{delK},\ref{rhos} to obtain a relation
between $\rho_s$ and the change $\Delta K$, 
in the $\omega>0$ spectral weight between
$T >T_c$ and $T \rightarrow 0$, finding

\begin{equation}
\frac{\rho_{s,observed}}{\Delta K} = 
\frac{2 \alpha}
{1+\alpha - 
(\rho_s(\Omega_+,\alpha=1)/\rho_s(\Omega_-,\alpha=1))}
\label{change}
\end{equation}
In this equation, $\rho_{s,observed}$ is the experimental $T\rightarrow 0$
value, $\rho_s(\Omega,\alpha=1)$ is the function shown in Fig. 5,
and the equation only applies if the gap $\Delta$ is well formed
and larger than $T$ for temperatures just above the resistive transition.

If $\Omega_+ < \Omega_-$ (as has been claimed \cite{Norman98a,Norman98b}
to occur in cuprates), then the ratio
is less than 2, if $\alpha <1$.  As shown in
\cite{Ioffe99} the ratio becomes equal to $2$ if $\Omega_+=\Omega_-$,
and could become greater than $2$ if the inequality were reversed.
The physics of the ratio is most easily undestood from the limit
$\alpha \rightarrow 0$, $\Omega_+=0$ and $\Omega_- >> \Delta$.  In
this case turning on an $\Omega >> \Delta$ strongly suppresses the 
conductivity in the region $\omega \sim \Delta$ but the small value
of $\alpha$ means the compensating increase in $\rho_s$ is
negligible. The values $\Omega_+ \approx 0$ and $\Omega_- \approx \Delta$
inferred from photoemission \cite{Norman98a,Chubukov98} lead to
$\rho_s/\Delta K =2 \alpha /(\alpha+0.4)$, i.e. to a substantial reduction
of the ratio from $2$ if $\alpha \sim 0.5$ or less.

Care is needed in applying Eq. \ref{change} to data.  As is clear from Fig. 3,
variations in $\Omega$ lead to variations in $\sigma$ over a wide energy
range, of order the basic scattering rate $\Gamma$.  An integration over
a smaller range could miss some contributions to $\Delta K$, leading
to a larger apparent value of the ratio.

\section{Comparison to Data and to Other Theories}

We begin by summarizing the results
obtained in the previous sections.  We used second order
perturbation theory in the interplane coupling to study the
interplane conductivity, with particular emphasis on the
effect of  phase fluctuations in the presence of a 
pairing gap $\Delta$.  We found that the formal consequence of
the existence of phase fluctuations is a Debye-Waller factor $\alpha < 1$
which multiplies the anomalous ('F') propagators.
The physical consequences include
a decrease in the total f-sum-rule oscillator strength,
a decrease in the superfluid stiffness $\rho_s$
and a change in the form of the conductivity at frequencies
of the order of twice the superconducting gap. Another important
parameter is $\Omega$ which is defined in Eq. \ref{Z} and
parametrizes the frequency scale associated with the mechanism
by which electrons are scattered.

As already noted in \cite{Ioffe99} the
most important and model-independent result
of this analysis 
is that the fundamental measure of the strength of quantum
fluctuations in the superconducting 
ground state is the ratio of $\rho_s$ to the area missing in $%
\sigma(\omega>0)$ as $T$ is decreased from above the  temperature
$T_{PG}$ at which the superconducting (or 'pseudo') gap becomes visible
to $T=0$. A $T=0$ $\rho_s$ which is smaller than the 
'missing area' implies non-negligible quantum fluctuations.

We further found that if quantum fluctuations are strong, 
the value of the c-axis conductivity in
the region $\omega \sim 2\Delta$ is increased above the predictions of BCS
theory. 
If the quasiparticles with $\omega \sim \Delta$ are weakly damped
(as indicated by photoemision experiments \cite{Shen95,Norman98b}) 
this increase takes the form of a peak. 

Finally, we derived a relation between $\rho_s$
and the area $\Delta K$ lost between $T=T_c$ and $T=0$ from the 
$\omega >0$ $\sigma$.  In the usual BCS
theory, where the gap closes at $T_c$, $\rho_s/\Delta K =1$.
If (as is believed to be the case in
underdoped curpates) a pairing gap exists exists in a wide
temperature regime above $T_c$ and the resistive transition
signals only the onset of phase coherence, 
then the ratio of $\rho_s$ to the weight $\Delta K$ lost
below $T_c$ is different from 1 and as seen
from Eq. \ref{change} depends on both the strength of the
quantum fluctuations and the variation with temperature of the
'offset parameter' $\Omega$.  If $\Omega$ is $T$-independent,
then the ratio is $2$ independent of $\alpha$, whereas if
$\Omega$ increases as $T$ is decreased below $T_c$ the 
ratio is less than 2, and if quantum fluctuations
are sufficiently strong can be less even than $1$.
For the values $\Omega(T>T_c) \approx 0$
and $\Omega(T \rightarrow 0) \approx \Delta$  inferred from
photoemission data \cite{Norman98a,Chubukov98,Shen95,Norman98b}
the ratio would become $1$ at $\alpha=0.4$.
The c-axis optics therefore 
contains information about the T-dependence of the electron 
scattering mechanism.
However, to obtain this information one must integrate the conductivity
over a wide range because we found changes in $\Omega$ led to
changes in $\sigma$ which extended over a  range of order 
the basic scattering
rate $\Gamma$.

Our results were obtained using second order
perturbation theory in the interplane hopping.  
This perturbation theory has been used by many workers
\cite{Anderson}, and in the present model may be tested.  The theoretical 
expansion parameter is $N_0t_{\perp}^2/\Gamma$, where $N_0$ is an
in-plane density of states, $t_{\perp}$ is an average of the hopping
over the fermi surface, and $\Gamma$ is the scattering rate.
The band structure estimates $t_{\perp}(p)=t_0(cos(p_x)-cos(p_y))^2$,
$t_0 \sim 0.15eV$ (for YBCO; smaller for others) and $N_0 \sim 2 states/eV$
\cite{Andersen94} lead to $N_0t_{\perp}^2 \sim 0.02eV$; the
$\Gamma$ inferred from the high frequency data on $YBa_2Cu_3O_{6.95}$
\cite{Homes95} then suggests that perturbation theory is very good.
One may make a more experimental estimate by writing (with units restored)
our theoretical result for the c-axis dc conductivity 
$\sigma_{c}^{dc}=\frac{e^2 d}{a^2}N_0t_{\perp}^2/\Gamma$.  Here
$d$ is the mean interplane spacing and $a$ is the in-plane lattice constant.
The observed $\sigma^{dc}\sim 100-200 \Omega^{-1} cm^{-1}$ for optimally
doped YBCO then implies $N_0t_{\perp}^2 \sim 1/50$, 
reasonably consistent with the above estimates and justifiying
the use of second order perturbation theory.  For other materials 
the interplane conductivity is even smaller, so the perturbation theory
should be even better.

Our results also rely crucially on the assumption that the part of the
Hamiltonian involving motion of electrons between planes has the form
given in Eq. \ref{H}, i.e. is the usual band-theory form which involves
hopping of a real electron in a manner which conserves in-plane momentum.
Strong scattering the barrier region (i.e. non-momentum-conserving hopping)
\cite{Radtke95,Abrikosov96,Kim98} or 'occupation modulated hopping'
\cite{Hirsch92} would invalidate our results.  We will argue below that the
close correspondence between our results and data suggests that the usual
band theory form of the hopping is the correct one.

We turn now to the data, beginning with the temperature and doping
dependence of the spectral weight. In overdoped materials a small increase
in low-$\omega$ normal-state spectral weight occurs as $T$ is decreased
below room temperature (see, e.g. Fig. 3a of \cite{Tajima97}). Our results
(c.f. Fig. 2)
suggest that this may be compensated by a small decrease in conductivity
over a wide frequency range. In optimally doped 
materials spectral weight seems to be conserved as
a function of temperature .  
In  optimally doped and 
overdoped materials the superconducting gap closes at
$T_c$ and spectral weight in $\rho_s$ compensates for the
area lost below $T_c$ \cite{Puchkov96,Tajima97,Homes95,Basov99a}. 
Further, in optimally doped and overdoped materials
the conductivity in 
the superconducting state appears to have the
usual BCS form, rising smoothly from the gap edge and
being always less than the normal state $\sigma$ \cite{Tajima97}. 
This is consisent with our results
if in these materials quantum and thermal
fluctuations of the order parameter
are negligible.

The observed $\sigma_c$ for optimally doped and overdoped
materials provides evidence against the alternative explanation of the $c$-axis 
conductivity advanced in \cite{Radtke95,Abrikosov96,Kim98}. In 
these works, the weak
frequency dependence of $\sigma_c$ at $T>T_c$ is attributed to a strongly
momentum non-conserving interplane coupling $t_\perp(p,p^{\prime})$, due
physically to strong scattering in the interplane barrier layers, rather
than to a large value of an in-plane electron scattering rate $\Gamma$. This
view implies that $c$-axis conductivity is equivalent to a point contact
tunneling; if it were correct then at $T<T_c$ one would expect to observe
peaks in $\sigma_c$ at $\omega=2\Delta$, corresponding to the sharp peaks
observed in photoemission and in $c$-axis tunneling 
\cite{Shen95,Norman98b,Renner98}.
Further, the momentum mixing caused by this scattering
would decrease $\rho_s$ below the BCS 'missing area' value, as noted by
Kim \cite{Kim98}. We therefore believe the assumption 
that the tunnelling matrix
element always conserves momentum, and that the changes in conductivity
with doping are due to changes in the underlying physics of the $CuO_2$
planes, is correct. The data are also inconsistent with the model of
\cite{Hirsch92} which for optimally doped materials
would predict a $\rho_s(T=0)$ greater than the
area missing below $T_c$.

We now consider the evolution of the spectral weight as the doping
is decreased.  Underdoped materials exhibit a normal state
'pseudogap', which begins to be visible as the temperature is
decreased below a temperature $T_{PG}$, which increases
with decreasing doping. Spectral weight appears to
be independent of $T$ at $T>T_{PG}$, with one
exception: in $La_{1-x}Sr_xCuO_4$,  $\int \sigma_c(\omega) d\omega$
appears to have a strong temperature dependence at all measured frequencies
between room temperature and low $100K$.  This behavior may be due to the
LO-TO structural phase transition, which will change the numerical value of
the interplane coupling.  In any event, this nonconservation of
weight over a wide scale  seems to be peculiar to
the $La_{1-x}Sr_xCuO_4$ materials and not to be related to
superconductivity. 

As the temperature is reduced
below $T_{PG}$ the onset of the 'pseudogap' causes  a decrease
in total spectral weight \cite{Puchkov96,Tajima97,Basov99a}. 
As the temperature is further reduced
below $T_c$ the weight (including both $\rho_s$ and 
$\sigma(\omega >0)$) increases again, but the $T=0$ weight is
never greater than the $T> T_{PG}$ weight. 
This behavior is consistent with our results if
the pseudogap is the superconducting pairing gap, and 
$T_c$ corresponds to the onset of phase coherence.  By contrast,
if the 'pseudogap' were caused by a charge or spin density wave
fluctuation, then the methods we have used here would predict that
weight would be conserved even in the $T_{PG} > T >T_c$ fluctuation regime:
weight lost at low frequencies due to the opening of the gap would
be shifted for frequencies just above $2 \Delta_{PG}$.  Thus we believe
the c-axis optical data provides strong support to the hypothesis that
the 'pseudogap' is due to  superconducting pairing without
phase coherence \cite{Emery95}. We note also that the model
of \cite{Hirsch92} did not consider the pseudogap regime explicitly,
but the results  seem to imply a $\rho_s(T=0)$
greater than the weight lost below $T_{PG}$, again in contradiction
to the data.

In underdoped $YBa_2Cu_3O_{6+y}$, $\rho_s$ at $T=0$ is less than the area
lost below $T_{PG}$ \cite{Homes95}, implying strong quantum fluctuations;
in other materials there is as yet no evidence for strong quantum fluctuations.
In $YBa_2Cu_4O_8$ the evidence suggests $\rho_s$ equals the area lost
below $T_{PG}$.

The ratio of $\rho_s$ to the change in $\omega >0$ spectral 
weight  between $T > T_c$ and $T \rightarrow 0$ appears 
qualitatively consistent with the results presented here.
In optimally doped and overdoped materials the gap closing
coincides with $T_c$ and the expected  ratio of $1$ is 
apparently observed \cite{Tajima97,Homes95,Basov99b}.  As the doping
is decreased and the pseudogap becomes more apparent,
the measured ratio grows.  A number of underdoped
materials appear to exhibit a ratio of 2
\cite{Tajima97,Homes95,Basov99a}, while a very recent 
measurement \cite{Basov99b}
finds a ratio of $\approx 1.6$ in a slightly underdoped BSCCO
sample.  It is not yet clear whether the effects of the T-dependent offset
are visible in the spectral weight. 
An experimental study of the interrelationship of $\alpha (T=0)$
(as defined by the ratio of the observed $\rho_s$ to the area
missing below $T_{PG}$), the T-dependent offset (from photoemission)
and the ratio of $\rho_s$ to the area missing below $T_c$ would
be very valuable.

We now turn to the frequency dependence of $\sigma_c$.  The effects
discussed here seem clearly to be visible in the $\sigma_c(\omega, T)$
presented in Fig. 3 of Tajima \emph{et al} \cite{Tajima97},
although it should be noted that there are large uncertainties in the
experimental determination of the electronic contribution to $\sigma_c$
because the observed conductivity is dominated by phonon lines which must be
subtracted out. Also the YBCO family of materials studied by
Tajima et. al. may have (especially for optimally and overdoped
$YBa_2Cu_3O_{6+y}$ and for $YBa_2Cu_4O_8$) large contributions from
electronic states involving the $CuO_2$ chains, which are not included
in our theory. Additionally the YBCO materials have a bilayer
structure which may allow other excitations, including in particular
'optical' Josephson plasmons \cite{Shibata98,Grueninger99}. 
Further study of systems without chains 
and of single-plane systems is needed.

Consider first the optimally doped sample (Fig. 3b of ref \cite{Tajima97}). 
The  $T>T_c$ conductivity is small and has 
negligible frequency and temperature
dependence, consistent with a large, quasistatic T-independent scattering
rate $\Gamma$. (Ref \cite{Homes95} shows the conductivity of a similar
sample over a wide frequency range--frequency dependence is only
apparent for frequencies $> 0.5 eV$ confirming the large $\Gamma$). 
As $T$ is decreased below $T_c$, the opening of the
superconducting gap leads to a decrease in $\sigma$ for $\omega < 600 \;
cm^{-1}$. We believe ({\it c.f.} Fig. 4) that
these data are consistent with a maximum gap value $%
\Delta_{max} \approx 200 \; cm^{-1}$ and an
offset $\Omega \sim \Delta$. The presence of absorption at the
lowest measured frequencies and the rounded shape of $\sigma_c(\omega)$ are
presumably due to a combination of defects and the fact that in real
materials $\sigma_c$ is determined by an average over the Fermi surface of
an angle dependent energy gap and an angle dependent c-axis hopping: $%
\sigma_c \sim \int d \widehat{p} t_{\perp}^2 \Pi(\widehat{p})$
(of course, additional contributions from chain states are also
possible). The observed $\rho_s$ is found to compensate
for the 'missing area'.  Thus, in
optimally doped $YBa_2Cu_3O_7$ the $T$-dependence of the total
spectral weight, of $\rho_s$ and 
of the $\omega$ and $T$ dependence of the $\sigma$ are
consistent with the behavior expected for a BCS superconductor with
negligible quantum fluctuations at low $T$ and a negligible pseudogap above $%
T_c$.

As the doping level is decreased, the behavior changes. As
seen in Figs 3c and 3d of \cite{Tajima97}, a pseudogap
(suppression of the absorption at $\omega < 2 \Delta$) is visible for a
range of temperatures $T>T_c$. The rise of the conductivity at the pseudogap
edge seems rather abrupt, as expected from the results shown
in Fig. 2.  The total spectral weight (superfluid part
plus contribution from $\omega>0$) is less than the room temperature
spectral weight. Next, a peak appears in $\sigma$ at $\omega \sim 400-500 \;
cm^{-1}$; the peak is stronger in the $YBa_2Cu_3O_{6.6}$ sample than in the $%
YBa_2Cu_3O_{6.7}$ one, but the position does
not shift. Finally, there is a hint of the minimum at $\omega
\sim 600 \; cm^{-1}$ in the $O_{6.6}$ sample.

This behavior is qualitatively consistent with the theoretical curves shown
in Fig.~4. It is reasonable to assume that as doping is decreased, quantum
fluctuations increase; this is shown by the analysis of the optical weight
presented in our previous paper and consistent with the analysis of the Hall
conductivity presented in \cite{Geshkenbein97}. The larger peak in the $%
O_{6.6}$ material is consistent with the stronger quantum fluctuations
expected there. The peak at $\omega > 400 \; cm^{-1}$ and minimum at $600 \;
cm^{-1}$ are expected from the $2\Delta, 2\Delta+\Omega$ structure of the
self energy if the threshold scattering frequency $\Omega$ is $\approx
\Delta \approx 200 \; cm^{-1}$. Finally, we note that in the $%
O_{6.6} $ material the spectral weight in the peak is about $50\%$ of the
spectral weight lost below $2 \Delta$ in superconducting state as expected
if the quantum fluctuations are very strong. The theory predicts an
anticorrelation of the weight in the peak and in $\rho_S(T=0)$. To make the
comparison quantitative one should measure both in units of the weght lost
below $2\Delta$. The data needed to make this comparison are not available
at present.

As noted above, an alternative interpretation of the peak as a bilayer plasmon
has been presented \cite{Shibata98,Grueninger99}.  In support of our
interpretation we note that 
the strength of the feature seems correlated with the strength
of quantum fluctuations.  However, our results imply that the peak
should be visible in single-layer materials where to date no peak
has been reported.
Indeed, in high-$T_c$ materials apart from $YBa_2Cu_3O_{6+y}$
no clear peak has been observed in $\sigma_c$
below $T_c$ (except, perhaps in $YBa_2Cu_4O_8$ 
where a slight hint of a peak is visible).
The absence of a strong peak in this material
is consistent with the observation that the
optical spectral weight is the same at room temperature as it is at $T=0$,
implying quantum fluctuations are weak in $YBa_2Cu_4O_8$.
Further experimental investigation of these issues would be very
helpful.

\section{Conclusions and Open Problems}

In this paper we have shown how quantum 
and thermal fluctuations of the superconducting
order parameter affect the integrated spectral weight
and frequency dependence of the interplane conductivity.
We showed that the data are consistent with the notion
that the normal state pseudogap is due to superconducting
pairing without long range phase coherence, and imply
that in very underdoped materials, quantum fluctuations of
the phase of the superconducting order parameter are strong.
No further assumptions are required to account for the data.
In particular, while the increase in spectral weight observed below
$T_c$ in underdoped materials may be interpreted as a change
in c-axis kinetic energy on entering the superconducting
state \cite{Anderson,Basov99a,Basov99b}, this increase was shown
to be a simple consequence of the hypothesis mentioned above,
and therefore provides little additional
insight into the microscopics of the
normal and superconducting states of high-$T_c$ materials.

The crucial microscopic information which can be extracted from
$\sigma_c$ involves the mechanism by which
electrons are scattered within a single $CuO_2$ plane.  We have argued
above, following Anderson \cite{Anderson}, that the c-axis conductivity
is in effect a spectroscopy of the in-plane Green function and implies
that at least for momenta
near the $(\pi,0)$ points which dominte the c-axis conductivity the in-plane
Green function is characterized by a self energy which is large, imaginary
and only weakly frequency dependent, i.e by a large frequency independent
scattering rate.  The physical origin of the scattering is not at present
understood.  One possibility, adopted by many 
workers \cite{Norman98a,Chubukov98,Ioffe98}
is that the scattering rate may be thought of in a relatively conventional
way, as the scattering of a usual electron off of some fluctuation.
An alternative view, propounded by Anderson \cite{Anderson} is that
the c-axis conductivity reflects a fundamentally unconventional
('non-fermi-liquid') physics of the $CuO_2$ planes, which have reasonably
well defined excitations which however have very small overlap
with the conventional electron.

In this paper we studied in detail the consequences of
the more conventional picture, because it is well enough defined
to allow detailed calculation. Because our results depend mainly
on the shape of the electron spectral function near the $(0,\pi)$
points, it seems likely they would follow from a non-fermi-liquid
picture also.

Within the conventional picture, at least,
it is natural to assume that
if the superconducting order parameter exhibits strong quantum fluctuations
then electron-electron interactions are strong generally and therefore
make an important contribution to the scattering rate. If this is the
case, then it seems reasonable that this scattering will be affected
by the onset of superconductivity, and specifically
that the low frequency part of it will be suppressed \cite{Littlewood92}.  
The hypothesized suppression of the scattering rate seems to have been
confirmed for high-$T_c$ materials by photoemission data.  The data are
however inconsistent with theoretical expectation in a manner which
deserves further discussion and investigation.  The argument
for the suppression was this \cite{Littlewood92}: electron-electron scattering
involves the creation of a particle-hole pair.  In the superconducting 
state the density of states of these pairs is reduced for energies
less than $2 \Delta$; hence in the range $\Delta < \omega <3\Delta$
one would have weakly scattered quasiparticles.
In fact, photoemission \cite{Norman98b} suggests that the range of weak scattering
is $\Delta < \omega <\Delta+\Omega$ with $\Omega \sim \Delta$ not $2\Delta$.
Norman et. al. \cite{Norman98a}, from a phenomenological point of view,
and Chubukov and Morr \cite{Chubukov98}, from a calculation of a model of
electrons coupled to spin fluctuations, explained this as a strong coupling
effect:  electrons do not scatter of a pair, but off of a collective mode
of some kind.  In the gapped state, what amount to excitonic
effects reduce
the final state energy below the naive $2\Delta$ threshold.

In order to produce the observed effects, the coupling 
of electrons to the mode must be very strong, and
in the normal state the spectral weight in the mode must be concentrated
at $\omega=0$.  This is already somewhat unusual, but the data exhibit
a further anomalous feature.  In underdoped materials the gap appears
at a high temperature $ T_{PG} \sim 200K$ while phase coherence appears
at the much lower resistive superconducting transition temperature
$T_c \sim 60K$.  {\it The appearance of the scattering rate
 offset, $\Omega$, seems to be tied to the onset of phase coherence,
and not to the opening of the gap}. This feature is not expected
from the arguments given above, and is not understood at present.

We have shown that, within the conventional picture at least, the T-dependent
offset leads to two effects in $\sigma_c$:  the appearance of a peak
at $\omega \sim 2\Delta$, if quantum fluctuations are strong, and a decrease
in the ratio between $\rho_s$ and the change between $T_c$ and $T=0$
in the $\omega >0$ pectral weight. The experimental status of these
two effects is unclear; further studies would be valuable.

On the theoretical side,
two possible avenues of investigation present themselves.  One is
that the mode which scatters electrons is the phase fluctuations of
the superconducting order parameter.  This idea was advanced by
Geshkenbein et. al. \cite{Geshkenbein97} and has been adopted by
us and by others \cite{Ioffe98,Franz98,Dorsey99}.  Another possibility
is that the more or less conventional physical picture of usual electrons
strongly scattered by some bosonic mode is simply inadequate, and that
the explanation should be sought in the physics of an underlying 
non-fermi-liquid state which becomes more conventional when long ranged
phase coherence is established.  This idea has been advanced by 
P. W. Anderson \cite{Anderson} and receives at least qualitative support
from the slave-boson gauge theory approach to the t-J model
\cite{Ioffe89,Lee89},
where establishment of phase coherence is related to the restoration
of more fermi-liquid-like behavior.

In our opinion, understanding the physics of the very large scattering
is one of the key problems in high-$T_c$ superconductivity. 
In this paper we have shown how measurements of the c-axis conductivity
can provide insights into this problem.

\textit{Acknowledgements} We thank D. Basov, S. Chakarvarty and
D. van der Marel and P. Pincus for helpful discussions. AJM
thanks NSF-DMR-9996282 for support, and the Aspen Center for
Physics for hospitality.

\end{multicols}

%\newpage

%\vspace{0.25cm} %\centerline{\epsfxsize=15cm \epsfbox{fig6.eps}} 

%{\footnotesize \textbf{Fig. 6}.  %Optical conductivity data from
%Tajima (ref \cite{Tajima97}.

% data from \cite{Tajima97}%


\begin{thebibliography}{99}

\bibitem{Ioffe99}  L.B. Ioffe and A.J. Millis, Science {\bf 285}
1241-44 (1999)

\bibitem{Puchkov96}  A. V. Puchkov, D. N. Basov and T. Timusk, J. Phys.:
Condens. Matter \textbf{8}, 10049 (1996).

\bibitem{Tajima97}  S. Tajima \textit{et al}, Phys. Rev. B \textbf{55}, 6051
(1997).

\bibitem{Anderson}P. W. Anderson, Phys. Rev. Lett. {\bf 67} 3844 (1991),
J. M. Wheatley, T. C. Hsu and P. W. Anderson, Phys. Rev. 
{\bf B37} 5897 (1988),
and P. W. Anderson, {\bf The Theory of Superconductivity
in the High-$T_c$ Cuprates}, Princeton University Press (Princeton, NJ:
1997); see especially pps 69-74.

\bibitem{Chakravarty94} S. Chakravarty and P. W. Anderson,
Phys. Rev. Lett. {\bf 72} 3859 (1994).

\bibitem{Chakravarty93} S. Chakravarty et. al. Science {\bf 261}
337 (1993).

\bibitem{Andersen94}  O. K. Andersen, O. Jepsen, A. I. Liechtenstein and I.
I. Mazin, Phys. Rev. \textbf{B49} 45-57 (1994) and O. K. Andersen,A. I.
Liechtenstein, O. Jepsen, and F. Paulsen, J. Phys. Chem. Sol. \textbf{56}
1573-92 (1995).

\bibitem{Schreiffer96}  J. R. Schreiffer {\it Theory of Superconductivity},
Addison Wesley (Reading, MA: 1983).

\bibitem{Radtke95}  R. J. Radtke and K. Levin, Physica \textbf{C250} 282
(1995).

\bibitem{Abrikosov96} A. A. Abrikosov, Phys. Rev. \textbf{B54} 12003 (1996).

\bibitem{Kim98}  E. H. Kim, Phys. Rev. \textbf{B58} 4252 (1998).
\bibitem{Norman98a}  M. R. Norman and H. Ding, Phys. Rev. B, \textbf{57},
R11089 (1998).

\bibitem{Chubukov98}  A. V. Chubukov, Europhysics Letters, \textbf{44}, 655
(1998); A. V. Chubukov and D. K. Morr, Phys. Rev. Lett., \textbf{81}, 471
(1998).

\bibitem{Littlewood92}  P. B. Littlewood and C. M. Varma, Phys. Rev. 
\textbf{B46} 405-420 (1992).

\bibitem{Homes95} C. C. Homes, T. Timusk, D. A. Bonn, R. Liang and W. N.
Hardy, Physica {\bf C254} 265 (1995).

\bibitem{Shen95}  Z. X. Shen and D. Dessau, Physics Reports \textbf{253}, 1
(1995).

\bibitem{Norman98b}  M. R. Norman et. al., \textit{et al} Nature, 
\textbf{392}, 157 (1998).

\bibitem{Emery95}  V. Emery and S. Kivelson, Nature \textbf{374} 434 (1995).

\bibitem{Corson99} J. Corson, R. Mallozzi, J. Orenstein, J. N. Eckstein and
I. Bozovic, Nature {\bf 398} 221 (1999).

\bibitem{Millis99} For a qualitative discussion see e.g. A. J. Millis,
Nature {\bf 398} 193 (1999)

\bibitem{Maldague77} P. F. Maldague, Phys. Rev. {\bf B16} 2437 (1977).

\bibitem{Baerswyl87} D. Baerswyl, C. Gros and T. M. Rice, Phys. Rev.
{\bf B35} 8391 (1987)

\bibitem{Millis90} A. J. Millis and S. N. Coppersmith, Phys. Rev. {\bf B42}
10807 (1990).


\bibitem{Hirsch92} J. E. Hirsch, Physica {\bf C199} 305-310 (1992).

\bibitem{Basov99a}  D. N. Basov et. al., Science \textbf{283} 49 (1999).

\bibitem{Renner98} Ch. Renner, B. Revaz, J.-Y. Genoud, K. Kadowaki and 
O. Fischer, Phys. Rev. Lett. {\bf 80} 1149 (1998).

\bibitem{Basov99b} A. S. Katz, S. I. Woods, E. J. Singley
T. W. Li, M. Xu, D. G. Hinks and D. N. Basov, unpublished
(cond-mat/9905170).

\bibitem{Shibata98} H. Shibata and T. Yamada, Phys. Rev. Lett. 
{\bf 81} 3519 (1998).

\bibitem{Grueninger99} M. Grueninger, D. van der Marel, A. A. Tsvetkov
and A. Erb, unpublished (cond-mat/9903352).

\bibitem{Ioffe98} L. B. Ioffe and A. J. Millis, Phys. Rev. {\bf B58},
11631-7 (1998).

\bibitem{Geshkenbein97}  V. B. Geshkenbein, L. B. Ioffe and A. I. Larkin,
Phys. Rev. B \textbf{55}, 3173 (1997).

\bibitem{Franz98}  M. Franz and A. J. Millis, Phys. Rev. \textbf{B58}%
14572-80 (1998).


\bibitem{Dorsey99} H.-J. Kwon and Alan T. Dorsey, Phys. Rev. {\bf B59} 6438
(1999).

\bibitem{Ioffe89} L. B. Ioffe and A. I. Larkin, Phys. Rev. {\bf B39}
8987 (1989).

\bibitem{Lee89} P. A. Lee, Phys. Rev. Lett. {\bf 63} 680 (1989) and
P. A. Lee, in {\bf High Temperature Superconductivity:  Proceedings}
K. S. Bedell, D. Coffey, D. E. Meltzer, D. Pines and J. R. Schreiffer,
eds.,, Addison Wesley (Reading, MA:  1990), p. 96 ff.


\end{thebibliography}
\end{document}